\documentclass{elsart}

\usepackage{natbib}
\usepackage{graphicx}

\usepackage{amssymb}

\begin{document}

\begin{frontmatter}



\title{Thermal emission from Isolated Neutron Stars and their surface 
magnetic field: going quadrupolar?}


\author[sil]{S. Zane,}
\author[rob]{R. Turolla}

\address[sil]{MSSL, University College of 
London, Holmbury St Mary, Surrey, RH5 6NT, UK} 
\address[rob]{Dept. of Physics, University
of Padua, via Marzolo 8, Padua, Italy}

\begin{abstract}

In the last few years considerable observational resources have been
devoted to study the thermal emission from isolated neutron stars. 
Detailed {\sl XMM} and {\sl Chandra} observations revealed a 
number of features in
the X-ray pulse profile, like asymmetry, energy dependence, and possible
evolution of the pulse profile over a time scale of months or years. 
Here we show that these
characteristics may be explained by a patchy surface temperature
distribution, which is expected if the magnetic field has a complex 
structure in which higher order multipoles contribute together with the 
dipole. We reconsider these effects from a theoretical point of view, and
discuss their implications to the observational properties of 
thermally emitting neutron stars.

\end{abstract}

\begin{keyword}
Neutron Stars \sep Magnetic Field \sep X-ray: Pulsations \sep Radiative 
Transfer 
\PACS 97.60.Jd	\sep 97.60.Gb 
\sep 97.10.Ex	\sep 97.10.Sj \sep 5.85.Nv	
\end{keyword}

\end{frontmatter}


\section{Introduction} 
\label{intro} The seven X-ray dim
isolated neutron stars (XDINSs) discovered so far (see e.g. Treves et 
al.~2000, Motch 2001 for a review) offer an unprecedented opportunity to
confront present models of neutron star (NS) surface emission with
observations. These objects play a key role in compact objects
astrophysics being the only sources in which we can have a clean view of 
the compact star surface. In particular, when pulsations and/or long
term variations are detected we can study the shape and evolution of the
pulse profile of the thermal emission and obtain information about the
thermal and magnetic map of the star surface. 
So far, X-ray pulsations have been detected in 
four XDINSs, with periods in the range 3-11 s. In each of 
the four 
cases the pulsed fraction is relatively large ($\sim 12 \%-35\%$) and, 
counter-intuitively, the 
softness ratio is maximum at the pulse maximum (Cropper et al.~2001, 
Haberl et al.~2003). Spectral lines have been 
detected in the soft X-rays and the line parameters may change 
with spin pulse. In addition, 
often the X-ray light curves appears to 
be slightly asymmetric and a gradual, long term evolution in 
both the X-ray spectrum and 
the pulse profile of the second most luminous source, RX J0720.4-3125 has 
been recently discovered (De Vries et al.~2004). 
\\
All these new findings 
represent a challenge for conventional atmospheric
models: the properties of the
observed pulse profiles (large pulsed fraction, skewness, and possibly 
time variations) cannot be explained by assuming that the thermal emission
originates at the NS surface if the thermal distribution is induced by a 
simple core-centered dipolar magnetic field. On the other hand, 
it has been realized
long ago that two effects can contribute at least to achieve relatively
large pulsed fraction (up to 20\%) : 1) radiative beaming (Pavlov et 
al.~1994) and 2) the presence of quadrupolar components in the magnetic field 
(Page and Sarmiento 1996). Here we present magnetized 
atmospheric models computed assuming a quadrupolar magnetic field geometry 
and show how they can
account for some of the general characteristics of the observed X-ray 
lightcurves (see Zane and Turolla 2004, in preparation for further 
details). 

\section{Getting to grips with the neutron star surface} 
\label{grips} 

\subsection{Computing the light curve} 
\label{mod} 

In order to compute the
phase-dependent spectrum emitted by a cooling NS as seen by a distant
observer, we basically perform three steps. First, we assume that the  
star  magnetic field possesses  a
core-centered dipole+quadrupole topology, $\mathbf B=\mathbf B_{dip}+
\mathbf B_{quad}$, where $\mathbf B_{quad} = \sum_{i=0}^4q_i\mathbf 
B_{quad}^{(i)}$.
The polar components of  $\mathbf B_{dip}$ and of the five
generating vectors $\mathbf B_{quad}^{(i)}$ are reported in 
Page and Sarmiento (1996). The NS surface temperature distribution is then 
computed using the simple  expression $
T_s = T_p\vert\cos\alpha\vert^{1/2}$, where $T_p$ is the polar 
temperature and $\alpha$ is the angle 
between the field and the radial direction, 
$\cos\alpha=\mathbf B\cdot\mathbf n$. Second, we compute the 
local spectrum emitted by each patch of the star surface by using 
fully ionized, magnetized hydrogen atmosphere models.\footnote{We 
caveat that partial 
ionization effects are not included in our work. Bound atoms and
molecules can affect the results, changing the radiation 
properties at relatively low $T$ and high $B$ (Potekhin et al.~2004).} 
Besides surface gravity, this 
depends on both the surface temperature $T_s$
and magnetic field, either its strength and orientation with
respect to the local normal. We introduce a
$(\theta,\, \phi)$ mesh which divides the star surface into a
given number of patches. 
The atmospheric structure and radiative transfer are then 
computed locally by approximating each atmospheric patch with a plane
parallel slab. Third, we collect the contributions of surface 
elements which are ``in view'' at different rotation phases (see Pechenick 
et al.~1983, Lloyd et al.~2003). 
We take the neutron star to be spherical (mass $M$, radius $R$) and
rotating with constant angular velocity $\omega = 2\pi/P$, where $P$ is
the period. Since XDINs are slow rotators ($P\approx 10$~s), we can
describe the space-time outside the NS in terms of the Schwarzschild
geometry. Under 
these assumptions, for a fixed polar temperature, 
dipolar field strength and surface gravity $M/R$, the computed
light curve depends on seven parameters: $b_i \equiv  q_i/ 
B_{dip}$ 
($i=0,...4$), the angle $\chi$ between line of sight and spin axis,  
and the angle $\xi$ between magnetic dipole and spin axis.

\subsection{Studying light curves as a population and fitting the 
observed pulse shapes}
\label{stat} 

Given this multidimensional dependence, 
an obvious question is whether or not we
can identify some possible combinations of the independent parameters that
are associated to particular features observed in the pulse shape.
Particularly promising for a quantitative classification
is a tool called principal
components analysis (PCA), which is
concerned with finding the minimum number of linearly independent 
variables $z_p$ (called the
principal components, PCs) needed to recreate a data set. 
In order to address this issue, we
divided the phase interval $(0 \leq \gamma \leq  1)$ in $32$ equally 
spaced bins,
and we computed a population of $78000$ light curves by varying $\chi, 
\xi$ and $b_i$ ($i=0,..4$). 
By applying a PCA, we found that each
light curve can be reproduced by using only the first $\sim 20$ more 
significant PCs and that the first four (three) PCs account for 85\% 
(72\%) of
the total variance. However, due to the 
strong non-linearity, we find so far difficult to relate the PCs 
to physical variables. \\
Nevertheless, the PCA can be regarded as an 
useful method to provide a 
link between the various light curves, since models ``close'' to each 
other in the PCs space have similar characteristics. From the PCA we 
obtain the matrix $C_{ij}$ such that $z_i 
\equiv C_{ij} y_j$, where $y_j$ is the observed X-ray intensity at phase 
$\gamma_j$ and $z_i$ is the $i$-th PC. Therefore, we can compute the PCs 
corresponding to each 
observed light curve and search the models population for the 
nearest solution in the PCs space (see fig.~\ref{fig1}, left). This 
in turn provides us a good trial solution, which can 
be used as a starting point for a 
numerical fit. The quadrupolar
components and viewing angles are treated as free parameters while the
polar values of $T_p$ and $B_{dip}$ are fixed and must fall in the domain 
spanned by our atmosphere model archive.\footnote{In general this will 
not contains the exact values inferred from spectral observations of 
XDINSs. However,
we have checked that a fit (albeit with different values of the
quadrupolar field) is possible for different combinations of $B_{dip}$ and 
$T_p$ in the
range of interest.}
Our preliminary results are illustrated in Fig.~\ref{fig1}, ~\ref{fig2}, 
and \ref{fig3}, and refer to $B_{dip} = 6 \times 10^{12}$~G, $\log T_p 
({\rm K}) =   
6.1-6.2$. 

\section{Summary of results} 
\label{conc} 

\begin{figure} 
\begin{minipage}{85mm}
\includegraphics[width=75mm,angle=0]{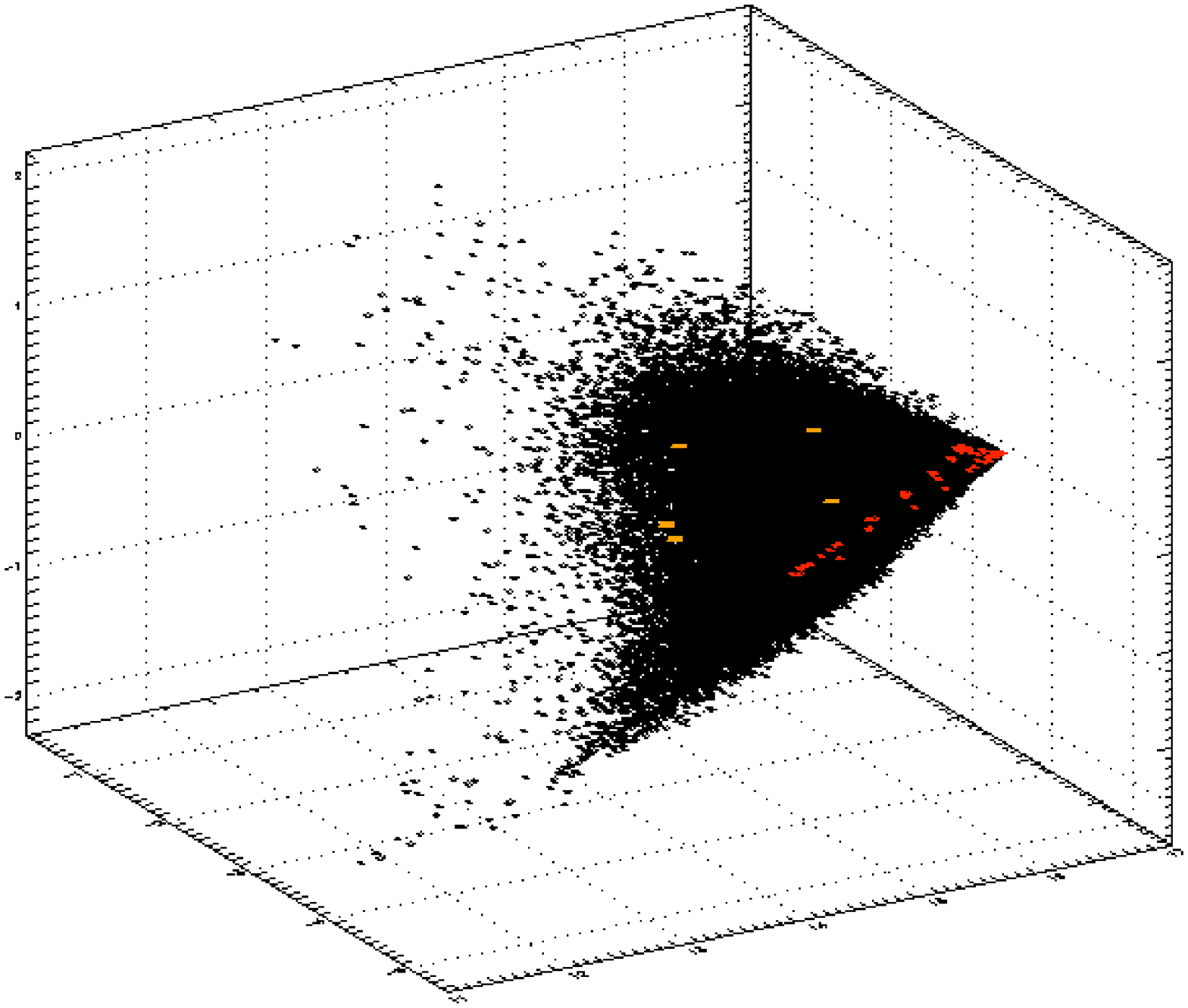}
\end{minipage}
\hfil\hspace{\fill}
%
\begin{minipage}{85mm}
\includegraphics[width=80mm,angle=0]{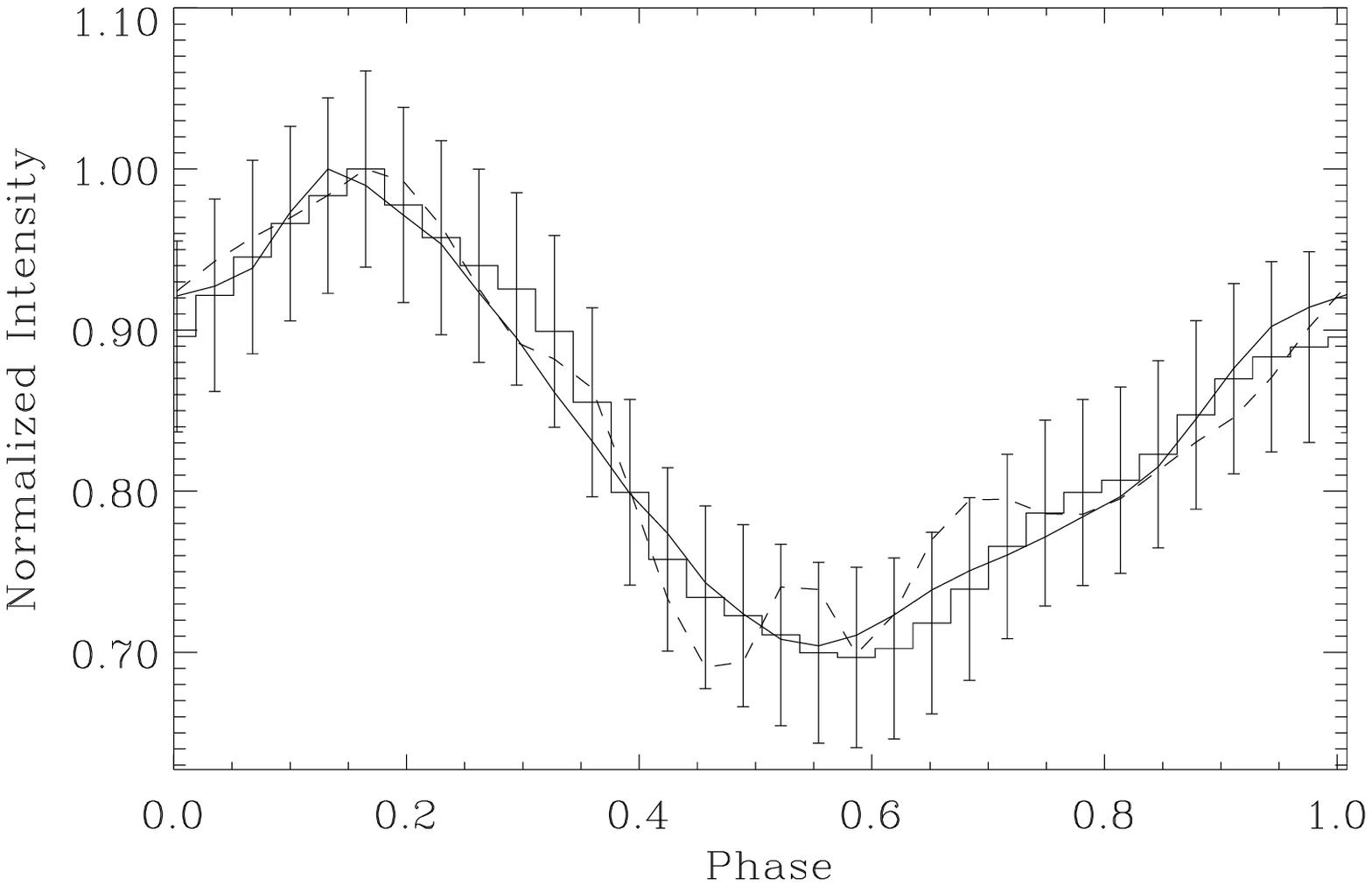} \end{minipage}
\caption{\label{fig1}}
\end{figure}

\begin{figure} 
\begin{minipage}{85mm}
\includegraphics[width=75mm,angle=0]{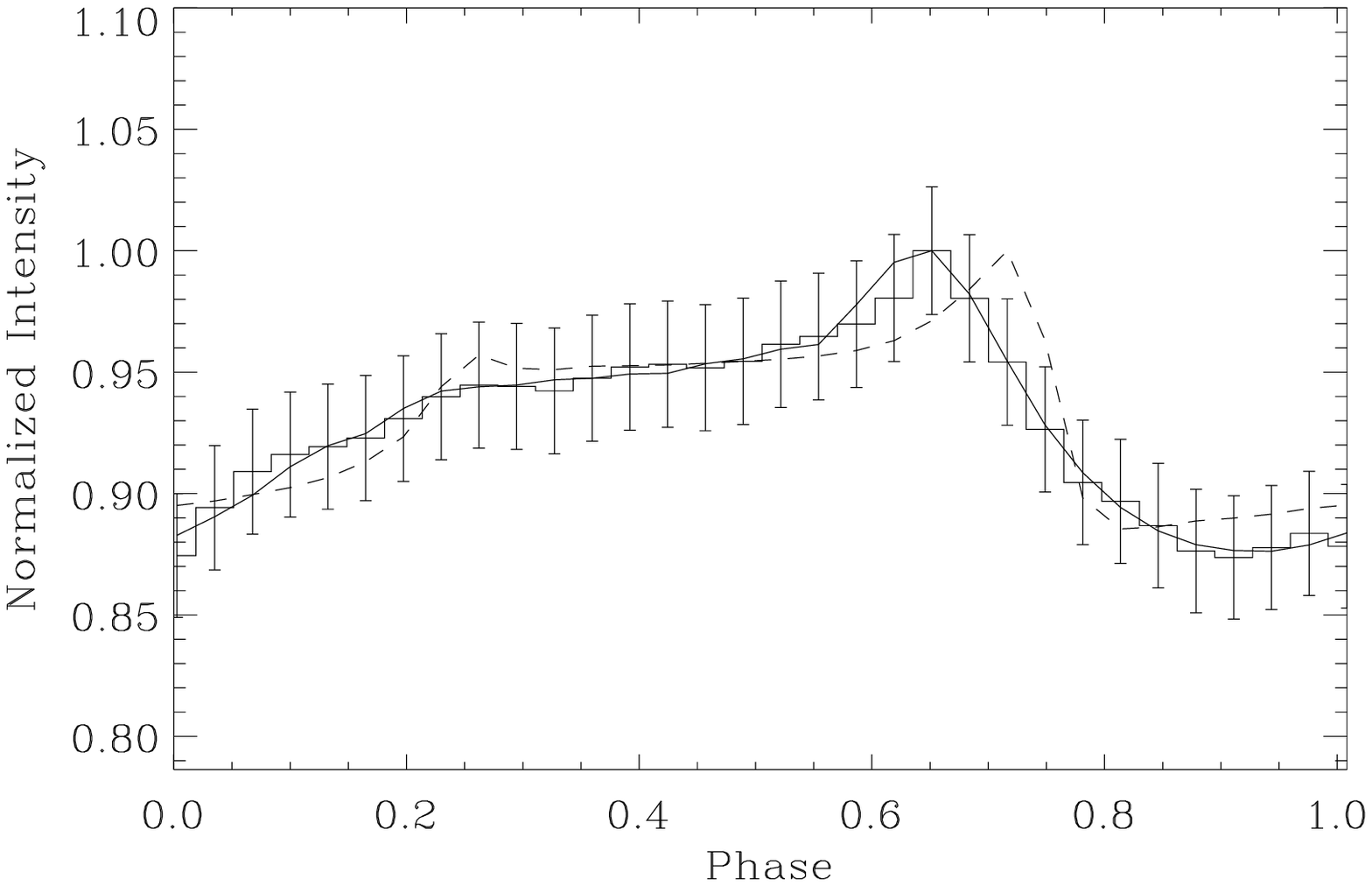}
\end{minipage}
\hfil\hspace{\fill}
%
\begin{minipage}{85mm}
\includegraphics[width=80mm,angle=0]{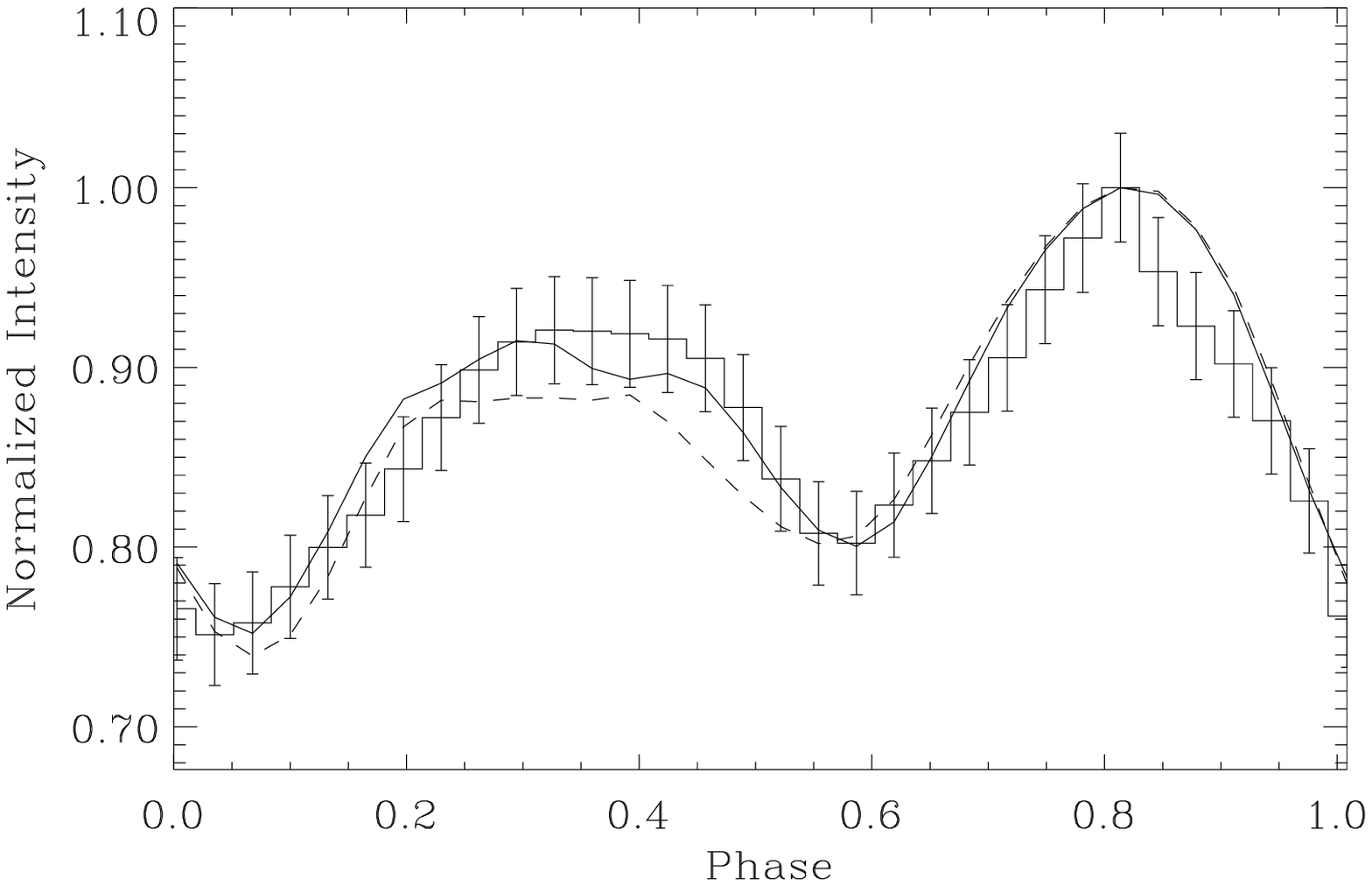} \end{minipage}
\caption{\label{fig2}} \end{figure}

\begin{figure} 
\begin{minipage}{85mm}
\includegraphics[width=75mm,angle=0]{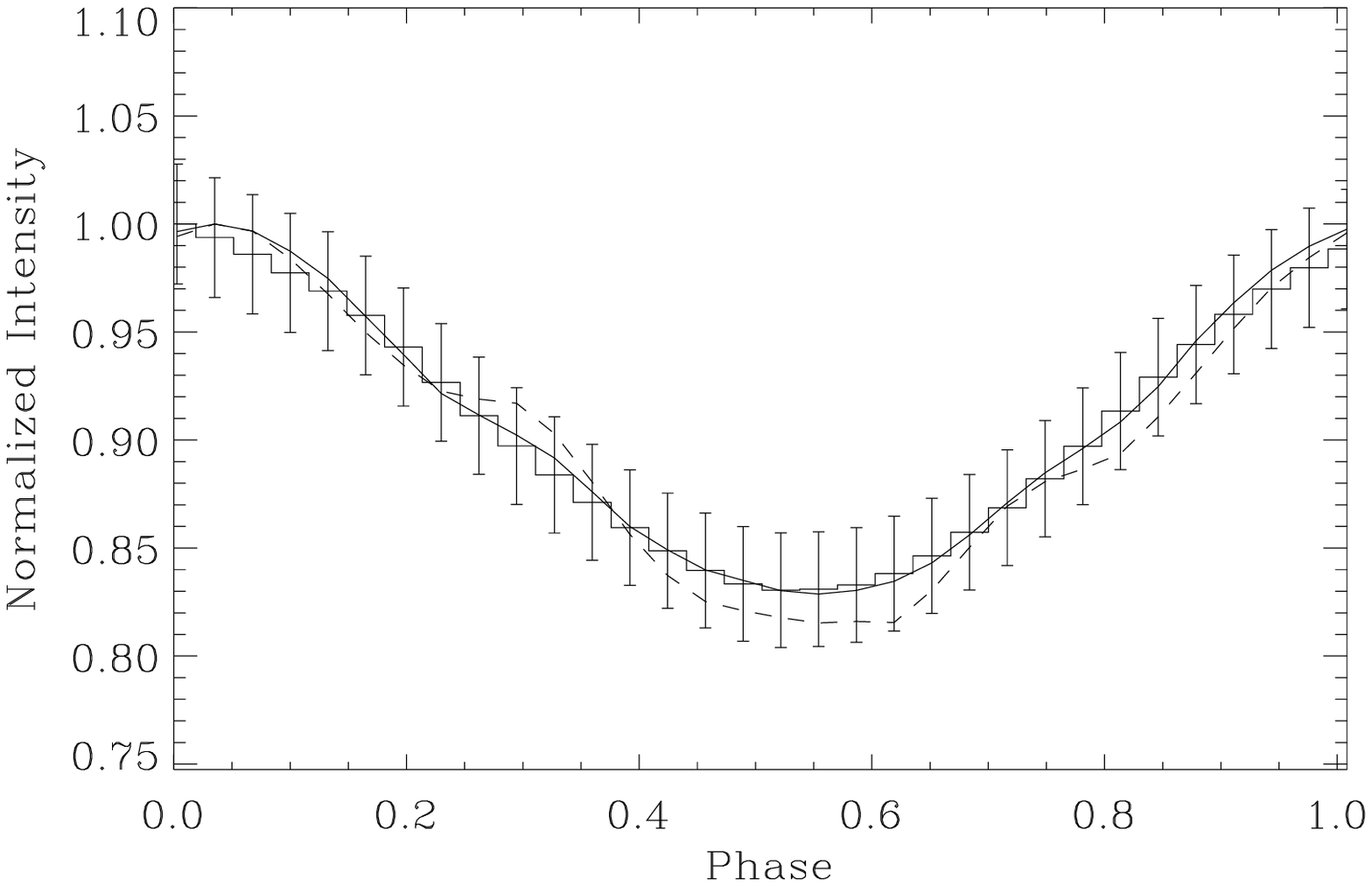}
\end{minipage}
\hfil\hspace{\fill}
%
\begin{minipage}{85mm}
\includegraphics[width=80mm,angle=0]{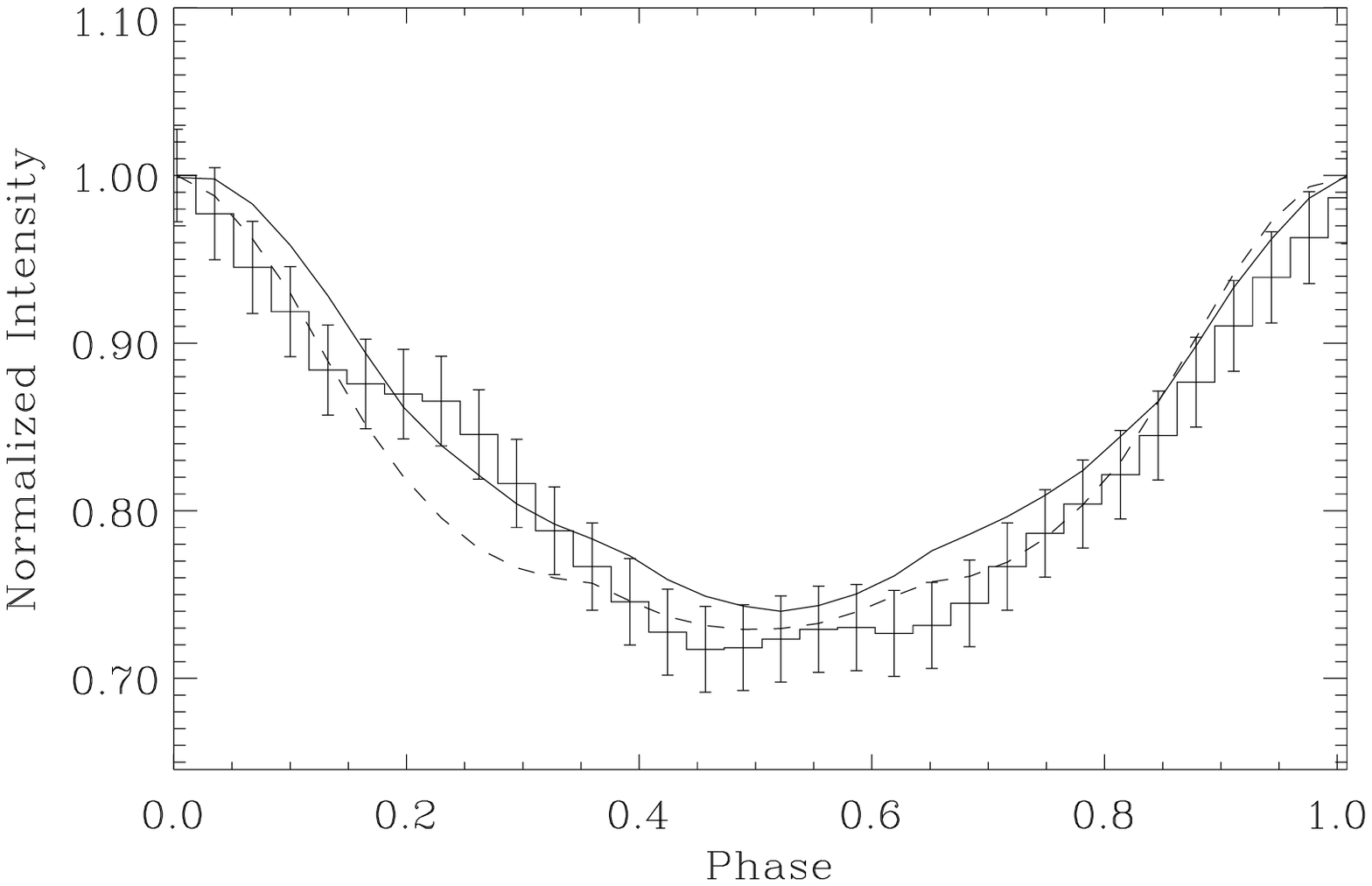} \end{minipage}
\caption{\label{fig3}} \end{figure}

As we can see from Fig.~\ref{fig1}, ~\ref{fig2} and ~\ref{fig3}, the 
broad characteristics of all the XDINS light curves observed so far 
are reproduced when allowing for a combination of quadrupolar 
magnetic field components and viewing angles. 
However, although in all cases a fit exists, we find that in general 
it may be not unique. This model has not a "predictive"
capacity in providing the exact values of the magnetic field components
and viewing angles: this is why we do not fit for all the parameters in a
proper sense and we 
do not derive parameter uncertainties or confidence levels. The goal is 
instead  
to show that there exist at
least one (and more probably more) combination of parameters that can
explain the observed pulse shapes, while this is not possible assuming a 
pure dipole configuration. 
In the case of RX J0720.4-3125, preliminary results show that the pulse 
variation 
observed between rev.78 and rev.711 can not be explained by a 
change in viewing angle only (as it should be if NS precession is 
invoked, de Vries et al.~2004) or 
magnetic field only. Instead, a change in all quantities (quadrupolar 
components and viewing angles) is 
needed.\footnote{How to produce field 
variations on such a short timescale needs to be addressed in more detail 
and, at present,
no definite conclusion can be drawn. For instance, a 
change of the magnetic field structure
and strength on a timescale of years may be conceivable if 
the surface field is small scaled (Geppert et al.~2003). 
In this case, even small changes in the inclination between line of 
sight and local magnetic field axis 
may cause significant differences in
the ``observed'' field strength.} 
Aim of our future work is to reduce the
degeneracy either by performing a more detailed statistical analysis on
the models population and  by refining the best fits solutions
by using information from the light curves observed in different color
bands and/or from the spectral line variations with spin pulse.




\begin{itemize} 

\item{Fig.~1. Left: The computed population of 78000
light curves plotted against the first 3 principal components. 
Axis are arbitrary; yellow   
symbols mark the position of the EPIC-PN light curves of the 
XDINSs (see next captions for references; both rev.~78 
and rev.~711 are shown for  RX J0720.4-3125), red symbols mark the light 
curves computed assuming a pure dipolar field. The PCs 
representation (limited to the first three PCs, which alone
account for the 72\% of the total variance) of the observed light curves
falls within the domain spanned by the quadrupolar model representations; 
this is why 
at least one fitting solution can be found. 
Right:  Fit of the EPIC-PN (0.12-0.7 keV) smoothed 
light curve of RX 
J0420.0-5022 (Haberl et al.~2004, rev.~570). Dashed line:  
trial 
solution as inferred from 
the closest model in the PCs space; solid line: best fit solution. The 
best fit parameters are: $b_0 = -0.48$, $b_1 = 0.02$, $b_2 = -0.25$, $b_3 
= 0.35$, $b_4 = -0.20 $, $\xi = 39.9^\circ$, $\chi=91.2^\circ $.}

\item{Fig.~2. Same as in the right panel of Fig.~\ref{fig1} for two
further INSs.  Left: Fit of the EPIC-PN (0.12-1.2 keV) light curve of RX
J0806.4-4123 (rev.~618, Haberl et al.~2004).  Here: $b_0 = 0.39$, $b_1 = 
-0.37$, $b_2 = 0.12$, $b_3 = -0.13$, $b_4 = 0.49$, $\xi = 0.0^\circ $, 
$\chi=59.2^\circ$.
Right: fit of the EPIC-PN (0.12-0.5 keV) light curve of RBS 1223 
(rev.~561, Haberl et al.~2003). Here: 
$b_0 = 0.21$, $b_1 = -0.02$, $b_2 = 0.59 $, $b_3 = 0.53 $, $b_4 = 0.50$, 
$\xi = 0.0^\circ$, $\chi= 95.1^\circ$.}

\item{Fig.~3. Same as in the right panel of Fig.~\ref{fig1} for 
the fit of two  
EPIC-PN (0.12-1.2 keV) light curves of  RX
J0720.4-3125 (both from De Vries et
al.~2004). Left 
(rev.~78): $b_0 = 
0.36$, $b_1 = 0.43$,
$b_2 = -0.16$, $b_3 = -0.16$, $b_4 = -0.39$, $\xi = 0.0^\circ $, 
$\chi=68.1^\circ$.
Right (rev.~711): $b_0 = 0.45$,
$b_1 = 0.49$, $b_2 = -0.06 $, $b_3 = -0.08 $, $b_4 = -0.26$, $\xi = 
0.0^\circ$,
$\chi= 87.6^\circ $.}

\end{itemize}

\clearpage 

\begin{thebibliography}{}


\harvarditem{Cropper et al.}{2001}{cr2001}
Cropper, M., Zane, S., Ramsay, G., et al. 
Modelling the spin pulse profile of the isolated neutron star RX 
J0720.4-3125 observed with XMM-Newton.  Astron. \& Astrophys., 365, 
L302-L307, 2001

\harvarditem{de Vries et al.}{2004}{dv2004}
de Vries, C.P., Vink, J., Mendez, M. et al. Long-term variability in the 
X-ray emission of RX J0720.4-3125.  Astron. \& Astrophys., 415, L31-L34, 
2004 

\harvarditem{Geppert et al.}{2003}{ge2003}
Geppert, U., Rheinhardt, M., Gil, J. Spot-like structures of neutron star 
surface magnetic fields. Astron. \& Astrophys., 412, L33-L36, 2003 


\harvarditem{Haberl et al.}{2003}{ha2003}
Haberl, F., Schwope, A.D., Hambaryan, V., et al. A broad absorption 
feature in the X-ray spectrum of the isolated neutron star RBS1223 (1RXS 
J130848.6+212708). Astron. \& Astrophys., 403, L19-L23, 2003 

\harvarditem{Haberl et al.}{2004}{hanoi2004}
Haberl, F., Motch, C., Zavlin, V.E., et al. The isolated neutron star 
X-ray pulsars RX J0420.0-5022 and RX J0806.4-4123: New X-ray and optical 
observations.  Astron. \& Astrophys., 424, 635-645, 2004 


\harvarditem{Lloyd et al.}{2003}{l03} 
Lloyd, D.A., Perna, R., Slane, P. et al. A pulsar-atmosphere model for PSR 
0656+14, astro-ph/0306235, 2003. 


\harvarditem{Motch}{2001}{mo01} Motch, C. 
Isolated neutron stars discovered by ROSAT, in: N.E.
White, G. Malaguti, and G.G.C. Palumbo (Eds.), 
X-ray Astronomy: Stellar 
Endpoints, AGN, and the Diffuse X-ray Background. 
NY: American Institute of Physics. AIP Conference Proceedings, Vol. 599, 
pp. 244-253, 2001.


\harvarditem{Pavlov et al.}{1994}{pa94} 
Pavlov, G.G., Shibanov, Yu.A.,  Ventura, J. et al. 
Model atmospheres and radiation of magnetic neutron stars:
Anisotropic thermal emission. Astron. \& Astrophys., 289, 837-845, 1994


\harvarditem{Page and Sarmiento}{1996}{ps96} 
Page, D., Sarmiento, A. Surface Temperature of a Magnetized Neutron 
Star and Interpretation of the ROSAT Data. II.  Astr. J., 473, 
1067-1078, 1996 


\harvarditem{Pechenick et al.}{1983}{p83} 
Pechenick, K.R., Ftaclas, C., Cohen, J.M. Hot spots on neutron stars - 
The near-field gravitational lens.  Astr. J., 274, 846-857, 1983 


\harvarditem{Potekhin et al.}{2004}{p04} 
Potekhin, A. Y., Lai, D., Chabrier, G., et al., Electromagnetic 
Polarization in Partially Ionized Plasmas with Strong Magnetic Fields and 
Neutron Star Atmosphere Models, ApJ, 612, 1034, 2004

\harvarditem{Treves}{2000}{tr00} Treves, A., Turolla, R., Zane, 
S., et al. Isolated Neutron Stars: Accretors and Coolers. PASP, 112, 
297-314, 2000 


\end{thebibliography}
\end{document}